\documentclass{llncs}

\usepackage{graphicx}
\usepackage{subfigure} 
\begin{document}
\bibliographystyle{splncs03}

\title{Extending search facilities via bibliometric-enhanced stratagems}

\author{Zeljko Carevic and Philipp Mayr}

\institute{GESIS - Leibniz Institute for the Social Sciences \\ Unter Sachsenhausen 6-8  \\ 50667 Cologne, Germany \\firstname.lastname@gesis.org}
\maketitle

\begin{abstract}

The paper introduces simple bibliometric-enhanced search facilities which are derived from the famous stratagems by Bates. Moves, tactics and stratagems are revisited from a Digital Library  perspective. The potential of extended versions of "journal run" or "citation search" for interactive information retrieval is outlined. The authors elaborate on the future implementation and evaluation of new bibliometric-enhanced search services. 

\end{abstract}

\section{Introduction}\label{sec:Introduction}
In information retrieval a change away from a mainly system-oriented perspective towards a more user-oriented perspective can be observed \cite{Ingwersen2005}. A main challenge is to gain insights into the way users perform searches in state-of-the-art Digital Libraries (DLs). During the past numerous models have been proposed that aim at modelling the information searching and seeking behaviour, e.g. the information seeking behaviour model proposed by Bates. According to Bates \cite{Bates1989} search activities can be separated into four categories: moves, tactics, stratagems and strategies. A move is a simple search activity like entering a query term or selecting a specific document. Tactics are a combination of many moves like the selection of a broader search term or breaking complex search queries into sub-problems.  Bates defines a stratagem as follows: "a stratagem is a complex of a number of moves and/or tactics, and generally involves both a particular identified information search domain anticipated to be productive by the searcher, and a mode of tackling the particular file organization of that domain." A stratagem could be for instance a "journal run" or a "citation search". Finally, a strategy is a combination of moves, tactics and stratagems that satisfies an information need like for instance searching for related work in a specific research area. 
It is difficult to implement strategic support in an information system as strategies involve numerous moves, tactics and stratagems as well as experience gathered in the entire search process. Although some work has been invested in developing strategic support \cite{Fuhr2005} for our position paper we focus on moves, tactics and stratagems. 
In state-of-the-art DLs moves and tactics are widely supported. For our real life example, the DL sowiport\footnote{sowiport.gesis.org}, users are enabled to enter search terms  selected from a search term recommender and to select broader or narrower terms from a thesaurus etc. \cite{Hienert2015}. Support for advanced search activities on the stratagem or strategy level does not exist.
Though the "systems file organization" covers information needed for stratagem support like for instance journal and citation data, we are missing predefined stratagems on the search interface (see Section \ref{Motivation}). Users therefore need to perform search stratagems manually. This requires deep knowledge of the information structure and the formulation of complex queries \cite{Booth2013}. This may be adequate for expert or power users but can hardly be accomplished by novices. We think that stratagems are an essential part of complex search tasks that need to be supported on the user interface. We believe that DLs like sowiport can largely benefit from predefined stratagems like footnote chasing, journal runs and citation search (see Section \ref{bibfac}).

\section{Related work}
The following related work section brings together some papers which have been identified as key ideas of the approach outlined later in this paper. 

First of all we want to mention the concepts developed by Bates \cite{Bates1989,Bates1990} which received a lot of attention in information and computer science. Her concepts describe the mechanisms of search activities and tasks in a very generalized way, as an information seeking model. These concepts of specific search tactics in an evolving search have been implemented in an academic Web environment by Fuhr et al. \cite{Fuhr2002,Fuhr2005}, in the project Daffodil. Today many other state-of-the-art DLs like Web of Science or PubMed support the search tactics outlined in Bates. 
Another important aspect is the ongoing popularity of the cognitive approach in IR and the inclusion of different forms of context and interaction (e.g. in Interactive IR) which has been combined by Ingwersen and J\"arvelin \cite{Ingwersen2005}. As a consequence of their framework the authors postulate a shift away from laboratory IR with controlled settings and without direct user engagement towards a holistic user-oriented perspective on the search process.

Bibliometric techniques are not yet widely applied to enhance the retrieval processes in DLs, although they offer value-added effects for users \cite{Mutschke,Mayr2014}. The objective of the IRM project\footnote{http://www.gesis.org/en/research/external-funding-projects/archive/irm/} was to introduce and evaluate bibliometric value-added services for information retrieval within a heterogeneous DL environment. The authors \cite{Mutschke} have investigated the use of informetrics as a ranking feature in a retrieval system. They found that using informetrics can improve the retrieval quality. Though the results are promising this has only been evaluated using the classic Cranfield setting. Although a prototype\footnote{http://multiweb.gesis.org/irsa/IRMPrototype} has been implemented that supports predefined bibliometric stratagems the larger scenario with real users and interactions has never been evaluated.

\section{Motivation}\label{Motivation}	
In the following we develop two basic positions which try to support the approach outlined in section \ref{bibfac}. Position A aims at the missing support for predefined stratagems on the user interface. In position B we briefly explain context-preserving and context free moves and how these could be supported in information systems. \newline
\newline
\bf{Position A: Missing predefined stratagem support on the user interface} \rm \newline
Like most DLs sowiport supports basic search features like facets, filtering, different types of ranking etc. When comparing the available features with Bates's search activities it can be seen that all these functions belong to moves and tactics. Domain specific search activities that could be considered as a stratagem are not supported on the interface. Thus, they need to be composed and executed manually by the user.  This requires knowledge of the domain and the underlying data structure. In DLs for example users need to be aware that papers are often published in journals or that co-cited documents are often related to each other. One problem with stratagems is that they are only used by experts that have experience in the given domain. A novice searcher may not be aware of certain stratagems which may result in limiting his/her search activities to moves and tactics. 
We think that stratagems are an essential part of complex search tasks that need to be supported. For this purpose we define a set of predefined stratagems that we consider useful for each level of experience (see section \ref{bibfac}).  \newline
\bf{Open Question:} \rm Which stratagems can be supported? \newline
\newline 
\bf{Position B: Supporting context-preserving and context free moves} \rm  \newline
When browsing DLs we can distinguish between a) context-preserving and b) context free moves. A context is defined e.g. by a query or filter criteria. In context-preserved browsing the context remains and is transferred into the next move. One example for a context-preserved move is faceted browsing. A user enters a search term and is then provided with a ranked result set. He/She could now reduce the result set by selecting a facet item. In this case, both the initial query (context) and the facet item are combined to a new query.
Browsing without context (b) is typically a simple move that performs a certain action in a retrieval system without transferring the context into the next step. An example of a context-free move may be the selection of an author appearing in the result set. This results in a new query  where the retrieval system performs a new search looking for the selected author name. 
We believe that both context-preserving and context free moves should be supported by the system. A system should offer both functions to the user and let him/her decide based on the current task what is best for him/her. Users should be able to decide whether they want to perform a context free or a context-preserved move. \newline
\bf{Open Question:} \rm How can context-preserving and context free moves be supported on the interface? How can these functions be arranged for the user? 

\section{Bibliometric-enhanced stratagems}\label{bibfac}
Bibliometric-enhanced stratagems aim at facilitating domain specific search activities by applying bibliometric measures for re-ranking and/or rearranging DL-entities like documents, journals or authors. Stratagems as described in the previous section can be implemented in various ways. A journal run for example can be implemented in the most simple way by offering a list of issues ranked by the publication date. We think that the implementation of stratagems depends on the current task thus, making it necessary to offer various ways of performing a stratagem search. To this end we describe a preliminary approach for novel bibliometric-enhanced search facilities as an extension to basic stratagem support. In the following we discuss two implementations for stratagem support in a DL like sowiport: an extended journal run and an extended citation search. Both examples are described using a mockup showing how to implement bibliometric-enhanced search facilities and how to deal with context free and context-preserving stratagems and moves.

\subsection{Journal run} 
\begin{figure}[h]\caption{Extended journal run}\label{extendedJournalRun}
 \centering
\includegraphics[scale=0.5]{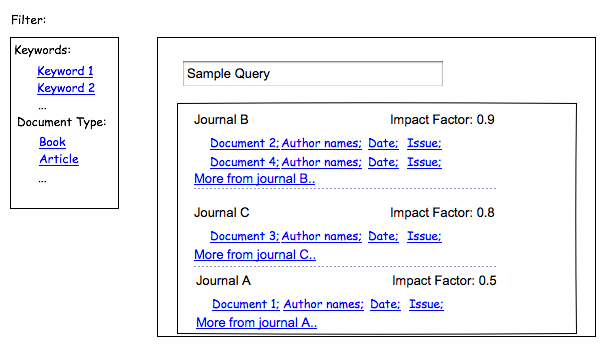}
    \end{figure}

A journal can be considered as a single specialized source for finding relevant documents from a manageable number of potential documents. On the other hand the focus on one journal results in the exclusion of other journals that might contain relevant documents. One way to overcome this issue is to offer different modes of a journal run on the user interface.
\begin{figure}[h]\caption{Journal Stratagem}\label{extended}
\includegraphics[width=\linewidth]{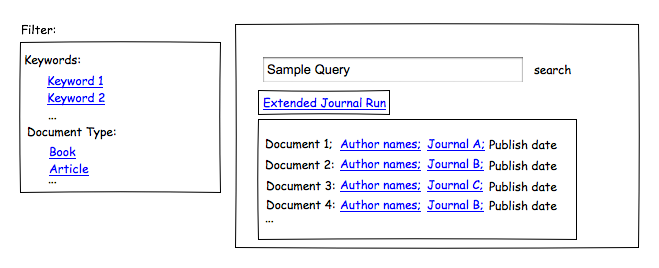}
\end{figure}
For this section two stratagems are described  (other modes are possible as well).\newline 1)  \it{Extended journal run}\rm : starting from a ranked result set (see Figure \ref{extended}) the user can perform an extended journal run that rearranges the articles based on the journal they were published in (see Figure \ref{extendedJournalRun}). It can be seen that an extended journal run changes the ranking from a document-based to a journal-based ranking. This journal-based ranking in our example is organized according to the impact factor measure of the journal. Using the impact factor is only one possible way of applying bibliometric journal metrics to re-rank the results. Other possible journal metrics are for example: h-index, g-index, etc. Each journal shows at least all documents that were available in the previous step. The list of documents can be expanded to all documents that are available for the particular journal by selecting "More from journal X".  \newline

2) \it{Context-preserving journal run}\rm : A context-preserving journal run is performed by selecting the name of a journal from a document appearing in the result set  (see Figure \ref{extended}).  In this example the previous moves and tactics that were performed before the journal run form the context of the stratagem. A context can be for instance a combination of the query term, a filter criterion and a single document attribute. Instead of ranking the documents in that journal by issue or by date we perform a ranking that is based on the context. Doing so we create a ranking that is orientated on the current search task. In comparison to the extended journal run this stratagem is limited to one journal.

\subsection{Citation analysis}
\begin{figure}[h]\caption{Citation Stratagem}\label{citations}
\includegraphics[width=\linewidth]{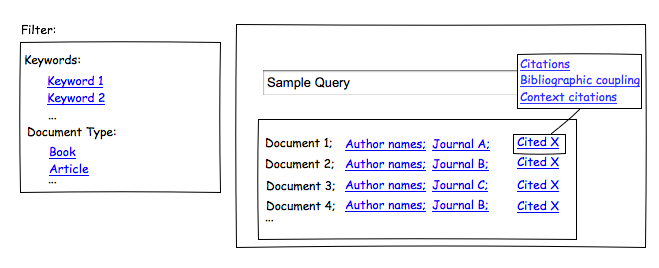}
\end{figure}

Another example for a bibliometric-enhanced stratagem is displayed in Figure \ref{citations}. 
In this example the user performs a citation search. We define a context menu from which the user can select different citation analysis modes. For this example three modes are proposed.\newline 1) A simple citation overview where the user can see a list of documents that cite the seed document. This is a simple move that performs a new search using the seed document as a query term looking for citations.\newline 2) The second mode allows the user to rank the citations based on bibliographic coupling \cite{Kessler1963}. Bibliographic coupling aims at finding related documents under the assumption that scientific papers are related to each other when they have one or more references in common. We now rank the citing documents based on the similarity in their reference lists. This way we expect documents that are strongly related to be ranked at the top of the list.\newline 3) The third mode is a context-preserving list of citations ranked by their relatedness to the seed. There are numerous methods of measuring the relatedness of the seed to the citing documents. The relatedness could for example be measured by comparing titles or by looking for keyword overlaps between the seed and the citing documents. This way citing documents that are related to the seed are ranked at the top of the list.

\section{Open Questions}
In this position paper we have described a preliminary approach for two novel bibliometric-enhanced search facilities as an extension to basic stratagem support. We strongly believe that bibliometric-enhanced search facilities can be a substantial part of DLs. Crucial points are: the choice of stratagems that could be supported and how these stratagems can be arranged on the interface. Furthermore, we need to investigate which bibliometric metrics can be integrated. 
One of the main challenges will be the evaluation of the stratagems. Our ideas for an evaluation go into two directions. We suggest a log-file based evaluation and a user evaluation. In the former we will measure the acceptance of the stratagems based on different indicators like session duration and positiv follow-up search activities (e.g. bookmarking or printing a document).  Additionally, we will create different A/B tests where a number of users are given some of the predefined stratagems instead of the current implementation. For our user evaluation we plan to conduct several user studies with experts. This way we want to gain an insight into which features could be helpful and which stratagems a system should support. 

\bibliography{relatedwork.bib}

\end{document}